\documentstyle[12pt]{ioplppt}

\begin{document}
\jl{6}
\letter{Order reductions of Lorentz-Dirac-like equations}

\author{J M Aguirregabiria, A Hern\'andez and M Rivas}

\address{F\'{\ii}sica Te\'orica, Universidad del Pa\'{\ii}s Vasco,
         Apdo.\ 644, 48080 Bilbao, Spain}

\begin{abstract}
We discuss the phenomenon of preacceleration in the light of a  method
of successive approximations used to construct the physical order
reduction of a large class of singular equations. A simple but
illustrative physical example is analyzed to get more insight into the
convergence properties of the method.
\end{abstract} 

\date{\today}

\maketitle

\section{Introduction}
\label{sec:intro}

In a recent paper~\cite{JMA} one of the present authors has proposed a
numerical implementation of a method of successive approximations that
allows automatic construction of the order reduction that contains the
physical, non-runaway,  solutions of a large class of singular
differential equations, including the classical equation of motion of
charged particles with radiation reaction~\cite{Rohrlich} and
fourth-order equations appearing in  theories of gravitation with a
quadratic lagrangian~\cite{Bel} and in the study of quantum corrections 
to Einstein equations~\cite{Parker}. Apart from its practical interest,
the convergence of the numerical method provides indirect but convincing
evidence of the convergence of the analytic method.

The goal of this letter is twofold: we want to discuss the phenomenon of
 preacceleration in the frame of the method of successive approximations
 and to produce a more physical exact example in which the method can be
 analyzed in full detail.

\section{Preacceleration}
\label{sec:preacc}

Among the puzzling properties of the Lorentz-Dirac
equation~\cite{Rohrlich},  which describes the motion of a radiating
point charge, the preacceleration  is one of the consequences of its
singular structure. Let us consider the non-relativistic approximation
to the Lorentz-Dirac equation (the so-called Abraham-Lorentz equation)
in the case of a charge $e$ that moves in a straight  line under the
action of an external force per unit mass $f(t)$:
 \begin{equation}
\ddot x=f(t)+\tau_0\,{\buildrel\ldots\over x},
\label{Abraham1}
 \end{equation}
where
 \begin{equation}
\tau_0\equiv\frac{2e^2}{3mc^3}.
 \label{tau0}
 \end{equation}
It is well known that if one eliminates the runaway solutions, the
physical motion is described by the integro-differential
equation~\cite{Plass,Rohrlich,JMA}
\begin{equation}
\ddot{x}=\int_0^\infty{\e^{-u}f(t+\tau_0u)\,du}.
\label{ALID1}
\end{equation}

If the external force is $f(t) = f_0\,\delta(t)$ it is enough to
insert this expression in equation~(\ref{ALID1}) to obtain the order
reduction
 \begin{equation}
\ddot x(t) = \cases{\displaystyle
  \frac{f_0}{\tau_0}\, \e^{t/\tau_0}&if $t<0$,  \\
  0 & if $t>0$. }
 \label{preacc}
 \end{equation}
According to this the charge would start accelerating before the pulse 
reaches it. 

This phenomenon has been widely discussed in connection with the
smallness of $\tau_0$ and taking into account the limitations of the
classical theory, but we want to analyze it here from the point of view
of the method of successive approximations~\cite{JMA}, which starts from
the approximation that neglects completely the radiation reaction
 \begin{equation}
 \ddot x=\Theta_0(t)\equiv f(t),
 \end{equation}
and iteratively constructs approximate reductions by substituting the
previous approximation on the right hand side of~(\ref{Abraham1}) 
 \begin{equation}
 \ddot x=\Theta_{n+1}(t)\equiv f(t)+\tau_0\,\Theta_n'(t)=
          \sum_{k=0}^{n+1}\tau_0^k\,f^{(k)}(t).
 \end{equation}
Under the appropriate mathematical conditions, this method will 
converge to the exact reduction
 \begin{equation}
 \ddot x=\Theta(t)=\sum_{k=0}^\infty\tau_0^k\,f^{(k)}(t),
 \label{ALser}
 \end{equation}
which is precisely the Taylor expansion of~(\ref{ALID1}). 

Now, one of the main hypothesis in the method of successive
approximations is Bhabha's remark~\cite{Bhabha} that the physical
solutions are precisely those that are regular in the limit 
$\tau_0\to0$, where according to~(\ref{ALID1}) and~(\ref{ALser}) one
recovers the second order equation
 \begin{equation}
 \ddot x=f(t).
 \label{radless}
 \end{equation}
But we can see that the preaccelerated solution~(\ref{preacc}) is
divergent in the limit $\tau_0\to0$, exactly as the remaining
pathological solutions. In consequence,  it cannot be constructed by the
(analytical or numerical) method of successive  approximations. One might
see this as a limitation of the latter method, but we think that it is
rather a limitation of the Abraham-Lorentz and similar equations.

To make clear our point of view we will consider a pulse of small but
non null width. For simplicity we will take a Gaussian pulse,
 \begin{equation}
f(t)=\frac{f_0}{\varepsilon\sqrt{\pi}}\,\e^{-(t/\varepsilon)^2},
 \label{gaussian}
 \end{equation}
but one could also consider any other pulse that recovers in the limit 
$\varepsilon\to0$ the value $f_0\,\delta(t)$. After
inserting~(\ref{gaussian}) in~(\ref{ALID1}) one gets the Newtonian
equation of motion that contains the  physical solutions:
 \begin{equation}
\ddot x=
\frac{f_0}{2\tau_0}\,\e^{t/\tau_0}\,\e^{\varepsilon^2/4\tau_0^2}\,
\mbox{erfc}\left(\frac{t}{\varepsilon}+\frac{\varepsilon}{2\tau_0}
\right).
 \end{equation}
This would be precisely the reduction constructed by the method of
successive approximations. By using the properties of the complementary
error function, it is easy to see that in the limit $\tau_0\to0$ one
recovers the radiationless result~(\ref{radless}), while for
$\varepsilon\to0$ one  gets the preaccelerated solution~(\ref{preacc}).
We see, thus, that these two limits do not commute. 

In our opinion the delta function obtained in the limit 
$\varepsilon\to0$ is beyond the field of applicability of Lorentz-Dirac
and  Abraham-Lorentz equations, for which one has to assume that the
applied force and acceleration do not change too much across a
time interval of length $\tau_0$, i.e. the radiation reaction cannot
be too important along such a tiny interval. Since this assumption is
not met by the delta function, this often useful limit is not applicable
here and one has necessarily to consider pulses of width larger than
$\tau_0$. This opinion is in agreement with the point of view of
authors~\cite{Bhabha,Valentini,Flanagan} that stress that the
analyticity with respect to $\tau_0$ is a fundamental hypothesis, which
is used in standard derivations of the Lorentz-Dirac equation.

\section{An exact example}
\label{sec:example}

In reference~\cite{JMA} we discussed a linear one-dimensional exact
example in the frame of the method of succesive approximations. 
We want to analyze now a three-dimensional exact example that,
though still linear, has a clearer physical meaning and will contribute
to our confidence on the convergence of the method of successive
approximations under appropriate conditions. Let us consider a charge
$e$ that moves in an external magnetic field $\bf B$, as happens in some
astrophysical contexts~\cite{Ginzburg} or in particle
accelerators~\cite{Schwinger}.
In the non-relativistic approximation the equation of motion is
 \begin{equation}
 \ddot{\bf x}={\bf\Omega}\times{\bf\dot x}+
 \tau_0{\bf \buildrel\ldots\over x},
 \label{Abraham}
 \end{equation}
where we have introduced the cyclotron frequency
 \begin{equation}
{\bf\Omega} = -\frac{e\bf B}{m},
 \end{equation}
which we will assume to be uniform and constant.
Starting from the lowest order
approximation
 \begin{equation}
 \ddot{\bf x}={\bf\Theta}_0\equiv{\bf\Omega}\times{\bf\dot x},
 \label{Abraham0}
 \end{equation}
we can construct successive approximations by using repeatedly
 \begin{equation}
 \ddot{\bf x}={\bf\Theta}_{n+1}\equiv{\bf\Omega}\times{\bf\dot x}+\tau_0
 \left[
 \frac{\partial{\bf\Theta}_n}{\partial t}+ 
 \left(\dot{\bf x}\cdot{\bf \nabla_x}\right) {\bf\Theta}_n+
 \left({\bf\Theta}_n\cdot{\bf \nabla_{\dot x}}\right) {\bf\Theta}_n
 \right].
 \label{Abrahamn}
 \end{equation}
It is straightforward to check that the successive approximations are 
 \begin{equation}
 {\bf\Theta}_{n}=\alpha_n\,{\bf\Omega}\times{\bf\dot x}-
 \beta_n\,{\bf\dot x}_\perp,
  \label{approxn}
 \end{equation}
where 
 \begin{equation}
{\bf\dot x}_\perp\equiv{\bf\dot x}-
\frac{{\bf\Omega}\cdot{\bf\dot x}}{\Omega^2}\,{\bf\Omega}
 \end{equation}
is the component of the velocity perpendicular to the magnetic field
and the constant coefficients are given by the recurrence
 \begin{eqnarray}
\alpha_{n+1}&=&1-2\,\tau_0\,\alpha_n\,\beta_n,\\
\beta_{n+1}&=&\tau_0\left(\Omega^2\alpha_n^2-\beta_n^2\right),
 \label{recurrence}
 \end{eqnarray}
and the initial conditions $\alpha_0=1$ and $\beta_0=0$. 

Recurrence~(\ref{recurrence}) has two fixed points 
$P_\pm=(\alpha_\pm,\beta_\pm)$ with
 \begin{eqnarray}
\alpha_\pm&=&\pm\frac{\sqrt{\frac{1}{2}
\left(\sqrt{1+16\tau_0^2\Omega^2}-1\right)}}{2\tau_0\Omega},\\
\beta_\pm&=&\frac{\pm\sqrt{\frac{1}{2}
\left(\sqrt{1+16\tau_0^2\Omega^2}+1\right)}-1}{2\tau_0},
 \end{eqnarray}
but a linear stability analysis proves that $P_-$ is always unstable and
that $P_+$ is asymptotically stable for
 \begin{equation}
 \tau_0\Omega < \frac{\sqrt{3+2\sqrt{3}}}{4}\approx 0.64.
 \label{range}
 \end{equation}
Furthermore, a simple bifurcation diagram in the dimensionless 
variables $(\alpha_n,\tau_0\beta_n)$ shows that the initial condition 
$(1,0)$ is in the basin of
attraction of $P_+$ and, in consequence, that the method of successive 
approximations
will in fact converge in the range~(\ref{range}) to the Newtonian equation
 \begin{equation}
 \ddot{\bf x}=\alpha_+\,{\bf\Omega}\times{\bf\dot x}-
 \beta_+\,{\bf\dot x}_\perp,
  \label{orderred}
 \end{equation}
which contains precisely the physical (non runaway) solutions 
for $\bf x$ found by
Plass~\cite{Plass}. Notice that all the approximations~(\ref{approxn}),
as well as the exact order reduction~(\ref{orderred}), are orthogonal to
the magnetic field, and that  the exact reduction exists even when the
method fails. This is not surprising because most approximation methods
have limited ranges of applicability. Moreover, in this case the method
will converge in all practical situations because the cyclotron
frequency is always very small compared to $1/\tau_0$. This simple but
illustrative exact example reinforces our conviction that the numerical
approximation  method~\cite{JMA} will converge in many cases of
interest.

\ack

This work has
been supported by The University of the Basque Country under contract
UPV/EHU 172.310-EB036/95.

\end{document}